\documentclass[pra,10pt,aps,showpacs,amsmath,amssymb,twocolumn]{revtex4-1}
\usepackage{graphicx}
\usepackage{dcolumn}
\usepackage{bm}

\usepackage{hyperref}

\begin{document}

\title{Magnetic liquids under high electric fields as broadband optical diodes}
\author{Jonas P. Pereira$^{1,2,3}$}
\email{jpereira@towson.edu, jonas.pereira@ufabc.edu.br}
\author{Igor I. Smolyaninov$^{4}$}
\email{smoly@umd.edu}
\author{Vera N. Smolyaninova$^{2}$}
\email{vsmolyaninova@towson.edu}

\affiliation{$^{1}$CAPES Foundation, Ministry of Education of Brazil, Bras\'ilia, Brazil}
\affiliation{$^{2}$Department of Physics, Astronomy and Geosciences, Towson University, 8000 York Road, Towson, Maryland 21252-0001, USA}
\affiliation{$^{3}$ Centro de Ci\^encias Naturais e Humanas, Universidade Federal do ABC, Av. dos Estados 5001, CEP 09210-580, Santo Andr\'e, SP, Brazil}
\affiliation{$^{4}$Department of Electrical and Computer Engineering, University of Maryland, College
Park, MD 20742, USA }
\date{\today}

\begin{abstract}
We show and give examples of how unidirectional propagation of light rays in the limit of geometric optics could arise in some magnetic fluids due to the magnetoelectric effect under weak DC magnetic fields and strong DC electric fields around half of their dielectric breakdown. For such liquids as kerosene and transformer oils, one-way propagation of light may occur for $30$ nm diameter magnetic nanoparticles (e.g. cobalt) and concentrations of $2\%$ or larger.
\end{abstract}

\maketitle

\section{Introduction}
As clear from its nature, the phenomenon of one-way propagation of light (an optical diode \footnote{In general, devices that allow light to propagate only in one direction are called optical isolators \citep{2013NaPho...7..579J}. We use the term ``diode'' instead of ``isolator'' in this work mainly because the media that we will investigate present strong nonlinear response to light propagation, allowing it to propagate only in one direction, besides having limiting features such as dielectric breakdown and specific DC electromagnetic fields for the effect to take place, which resemble some of the aspects conventional electronic diodes exhibit.}) requires
violations of time-reversal and parity symmetries of a physical system, leading the frequency ($\omega$) of a  propagating wave to behave differently for opposite directions of its wave vector $\vec{q}$, $\omega(\vec{q})\neq \omega(-\vec{q})$ (see \citep{2001PhRvE..63f6609F} and references therein).
This situation occurs naturally in the presence of external electromagnetic fields in magnetoelectric media \cite{2001PhRvE..63f6609F,landau1984electrodynamics}.
In order to render the (nonlinear) magnetoelectric effect experimentally noticeable (it is usually very small in magnitude and not common in material media \cite{2001PhRvE..63f6609F,landau1984electrodynamics}), one should work with metamaterials. In such engineered media the basic constituents are much smaller than the wavelengths of the propagating waves, resulting in controllable dielectric coefficients (permittivity and permeability) \citep{1968SvPhU..10..509V, PhysRevLett.84.4184,2001Sci...292...77S,2010opme.book.....C}. In this regard, it seems that magnetic liquids, host liquid media in the presence of magnetic nanoparticles \citep{2005RuCRv..74..489G}, could be potential candidates for optical diodes since their magnetoelectric response to external fields is controllable and can reach high values.

Magnetic nanoparticles are easily produced and controlled \citep{2005RuCRv..74..489G} and naturally lead to self-assembled three-dimensional metamaterials when coated (with fatty acids) and inserted into a host liquid medium (usually oils or kerosene) under external magnetic fields \citep{2014NatSR...4E5706S}. This situation has clear advantages  compared to  challenging nanofabrication of three-dimensional metamaterials, due to their intrinsic small dimensions compared to the light wavelengths.
Magnetic nanoparticles are also very attractive due to their use in several areas of biomedicine, such as cancer treatments and magnetic resonance imaging (see, e.g., \citep{2003JPhD...36R.198B,issa2013magnetic} and references therein), and in power lines for cooling transformers and enhancing their dielectric properties \citep{2005JMMM..289..415K,2006JOSAB..23..498W}.

Conventional optical diodes are usually related to photonic crystals (periodic structures whose basic constituents' dimensions are of the order of wavelengths of the propagating waves) by means of the magneto-optic effect;  see, e.g., \citep{wang2009observation, PhysRevB.83.075117,2013NatMa..12..175Y, PhysRevLett.100.023902,2013OptMa..35.1455F} and references therein. They can also be related to other plasmonic designs such as nonlinear chiral meta-atoms \citep{2011NJPh...13c3025S}, nonlinear chiral ultra-thin three-layered structures \citep{2012PhRvL.108u3905M}, structures constructed with nonlinear silicon microrings (without the presence of magnetic fields) \citep{2012Sci...335..447F}, periodically poled waveguides \citep{2001ApPhL..79..314G}, two-layered ultra-thin metamaterials (both metallic) \citep{li2014broadband}, etc. For further systems and effects that lead to diode-like devices (optical isolators), we refer the reader to Refs. \citep{2011NJPh...13c3025S,2012PhRvL.108u3905M,2012Sci...335..447F,2001ApPhL..79..314G,li2014broadband,2016OExpr..2415362Y}. The majority of diode-like devices works in the infrared and terahertz range, are limited in size and are not economically interesting to be scaled up \citep{2016OExpr..2415362Y}. Besides, nonlinear devices based on the Kerr-effect intrinsically have limitations in their nonreciprocity, which restrict their actuation as optical isolators  \citep{2015NaPho...9..388S}.

\begin{figure}[!htb]
 \centering
 \includegraphics[width=\columnwidth]{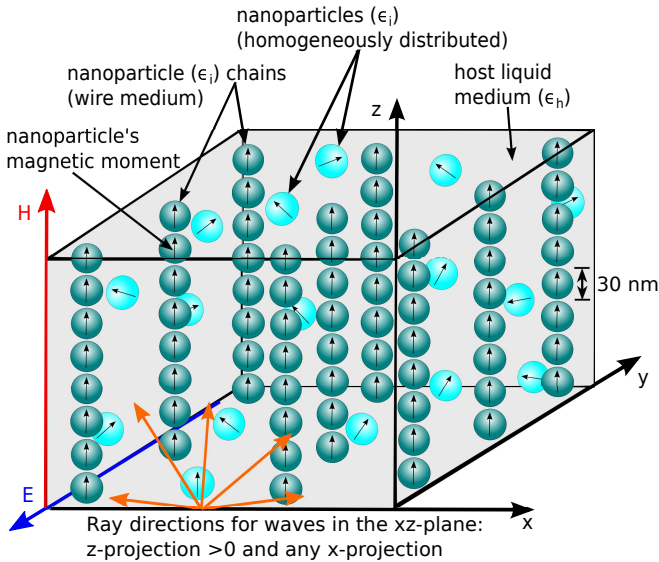}
\caption{{\small\sf (color online). Schematic representation of a magnetic liquid in the presence of external DC electric $E(-\hat{y})$ and magnetic $H\hat{z}$ fields and $30$ nm magnetic nanoparticles ($\epsilon_i$) in a host liquid medium $(\epsilon_h)$. Chained nanoparticles have their magnetic moments aligned not mainly due to the external magnetic field, but due to the fields of their nearest (coated) neighbors. Misaligned nanoparticles, which are the same in nature as the ones in the chains, have their magnetic moments well less aligned with the external magnetic field and are homogeneously distributed in the host liquid. Aligned nanoparticles give rise to a wire medium whose effective permittivity tensor is anisotropic and given by Eqs.~(\ref{e11typ}) and (\ref{e33typ}) (Section II). For waves propagating in the whole $xz$-plane, for instance, our system is such that the projection of rays along the $z$-direction is only positive, which shows its unidirectional propagation aspect (Section III).
}}\label{schematics}
\end{figure}

In this work we show that some magnetic liquids could also behave as optical diodes in the limit of geometric optics (so that this effect is intrinsically associated with light ray propagation) if subjected to convenient external fields and do not exhibit the above-mentioned problems.  More explicitly, we show that the optics of some liquids (transformer oils or kerosene) with magnetic nanoparticles (cobalt or magnetite) of around $30$ nm diameter and in concentrations around $2\%$ or larger by volume  in the presence of small applied DC magnetic fields and strong DC electric fields (around half the fluid's breakdown) may result in one-way propagation of light rays (see their schematics in Fig. \ref{schematics}), mainly due to the magnetoelectric effect they naturally exhibit (therefore, there is no limitation in their nonreciprocity). Given that magnetic fluids are self-assembled metamaterials already commercially available, optical diodes associated with them would be easily produced, inexpensive and large in size, which has clear advantages compared to the previously mentioned optical diode schemes.

The plan of this work is as follows. In the next section we revisit the effective dielectric properties of magnetic liquids in the presence of magnetic fields. Section III is devoted to the investigation of a model of an optical diode based on magnetic fluids subjected to high electric fields. In Section IV we summarize the main issues raised in this paper. We work with Cartesian coordinates and unless otherwise specified, the speed of light in the vacuum is taken as unity.

\section{Effective dielectric coefficients of magnetic liquids}

Whenever one considers inclusions in a host medium that line up in the presence of an external magnetic field \footnote{The external magnetic field triggers the effect; the lining up mainly occurs due to the magnetic fields of the nearest neighbor coated nanoparticles \citep{2014NatSR...4E5706S}.}, assumed for specificity to be in the $z$-direction, one comes up with an effective uniaxial medium whose principal dielectric coefficients may be calculated based on the Maxwell-Garnett's approximation as \citep{2010opme.book.....C,2006JOSAB..23..498W}
\begin{equation}
\epsilon_{11}=\epsilon_{22}=\frac{2\alpha \epsilon_i \epsilon_h +(1-\alpha)\epsilon_h(\epsilon_i+\epsilon_h)}{(1-\alpha)(\epsilon_i+\epsilon_h)+2\alpha \epsilon_h}\label{e11}
\end{equation}
and
\begin{equation}
\epsilon_{33}=\alpha \epsilon_i+ (1-\alpha)\epsilon_h\label{e33},
\end{equation}
where $\alpha$ is the volume fraction of the lined up inclusions, whose permittivity is $\epsilon_i$, and $\epsilon_h$ is the permittivity of the host medium. We will assume that in the frequency range of interest, namely the near-infrared, the effective medium in non-magnetic, $\mu=1$, and therefore one can take $\vec{B}$ or $\vec{H}$ for the magnetic field without distinction. Equations (\ref{e11}) and (\ref{e33}) are actually good approximations for the effective permittivity tensor of a composite medium only when $\alpha\ll1$.
In this case, one can Taylor-expand Eq.~ (\ref{e11}) up to first order in $\alpha$, yielding
\begin{equation}
\epsilon_{11}=\epsilon_{22}\approx \epsilon_{h} -2\alpha\epsilon_h\frac{\epsilon_h-\epsilon_i}{\epsilon_h+\epsilon_i}\label{e11aprox}.
\end{equation}

The alignment of the magnetic inclusions due to the presence of a magnetic field takes place when they are metallic, coated with fatty acid as a surfactant layer \citep{2004JMMM..272.2377K,2014NatSR...4E5706S} and have a nonzero resultant magnetic moment. In such a case we may assume that $\alpha=\alpha(T,H)$, where $T$ is the temperature of the system and $H$ the magnitude of the external magnetic field \citep{2014NatSR...4E5706S}. In this way, one ends up effectively with the permittivity being dependent upon the magnetic field. Besides, due to the metallic inclusions present, the effective permittivity tensor should also depend upon $\omega$.

Note that the total volume fraction in the host medium is a given constant, naturally related to very large values of the magnetic field (situation where all nanoparticles are aligned and participate in the aligned chains), and because of that it will be denoted by $\alpha_{\infty}$. This means that the quantity $(\alpha_{\infty} - \alpha(T,H))$ is the volume fraction of the inclusions misaligned with the magnetic field. They contribute to the permittivity of the host medium for a given $H$. Therefore, based on the Maxwell-Garnett aproximation for homogeneously distributed inclusions in three-dimensional space \citep{2010opme.book.....C}, one has to make the substitution $\epsilon_h\rightarrow \epsilon_h[1-3(\alpha_{\infty}-\alpha(T,H))]$ in Eqs.~(\ref{e11})--(\ref{e11aprox}).
In our subsequent analysis we will consider the typical case $|\epsilon_i|\gg \epsilon_{h}$, which naturally happens in the near-infrared (see, e.g., \citep{2014NatSR...4E5706S} and references therein). In such a case and taking into account the above-mentioned effect of misaligned nanoparticles in the host medium, we have (again up to first order in both $\alpha$ and $\alpha_{\infty}$)
\begin{equation}
\epsilon_{11}=\epsilon_{22}\approx \epsilon_h(1+3\alpha_{\infty}-\alpha)\label{e11typ}
\end{equation}
and
\begin{equation}
\epsilon_{33}\approx \epsilon_h-\alpha|\epsilon_i|\label{e33typ}.
\end{equation}
One sees from Eq. (\ref{e33typ}) that the case $\epsilon_{33}\approx 0$ may take place in some frequency range. There, though, $\epsilon_{11}$ is a given positive constant. It is important to keep in mind that its value is slightly different from $\epsilon_h$ and depends upon the external magnetic field through $\alpha$. As we will see in the next section, this is a fundamental criterium for the effect of one-way propagation of light, along with the presence of a suitable external electric field.

\section{A model for one-way propagation in magnetic fluids}

We now elaborate upon the properties that magnetic fluids (specifically, the ferrofluids) should have for behaving as optical diodes.
Let us assume that an external DC electric field is present in a direction perpendicular to the DC magnetic field, that we define as the $y$-direction, $\vec{E}=E\hat y$ ($E$ can be either positive or negative). For an uniaxial medium whose permittivity tensor has principal components $\epsilon_{11}, \epsilon_{11}$ and $\epsilon_{33}$ depending upon the magnetic field, one can show that in the limit of geometric optics its Fresnel equation \citep{1975ctf..book.....L,1999poet.book.....B} splits into (see the general formalism in \citep{2014PhRvA..89d3822D})
\begin{equation}
\epsilon_{11}\omega^2-\vec{q}^{\;2}+\frac{\partial \epsilon_{11}}{\partial B}E\omega q_{x}=0\label{disprelex}
\end{equation}
and
\begin{equation}
(\epsilon_{11}-\epsilon_{33})q_z^2 -\epsilon_{11}(\vec{q}^{\;2}-\omega^2\epsilon_{33})=0\label{disprelor},
\end{equation}
where $\vec{q}$ and $\omega$ are defined as the wave-vector and angular frequency of the propagating wave, respectively, $q_i$, $i=(x,y,z)$, is any of its Cartesian components, and we recall that we have assumed $\mu=1$.
One can verify that an electric field parallel to the magnetic field does not produce any effect on the wave propagation.
The solution to Eq. (\ref{disprelor}) can be cast as
\begin{equation}
v\doteq v_0^{\pm}=\pm\sqrt{\frac{\epsilon_{11}+(\epsilon_{33}-\epsilon_{11})\hat{q}_z^2}{\epsilon_{33}\epsilon_{11}}}\label{vphaseo},
\end{equation}
while that to Eq. (\ref{disprelex}) simply reads
\begin{equation}
v\doteq v_1^{\pm}=\frac{1}{2\epsilon_{11}}\left[-\frac{\partial \epsilon_{11}}{\partial B}E \hat{q}_{x}\pm \sqrt{4\epsilon_{11}+\left( \frac{\partial \epsilon_{11}}{\partial B}E \hat{q}_{x} \right)^2}\right]\label{vphaseex},
\end{equation}
where we have defined $\hat{q}_i\doteq q_i/|\vec{q}|$. We recollect that $\vec{v}= v\hat{q}$. From Eq. (\ref{vphaseo}) one clearly sees that opposite directions are totally equivalent and therefore it represents a sole physical solution. Though in the case of Eq. (\ref{vphaseex}) opposite directions are dissimilar, it also represents a single physical wave whenever $\epsilon_{11}>0$, which is the case that we will consider here. Therefore, waves would propagate in all directions. Notwithstanding, the important quantity in the limit of geometric optics is the ray and we should attach unidirectional propagation to it. We note that in such a limit it is meaningful to define rays as a superposition of waves even in the case of nonlinear media (see \citep{2013PhRvD..88f5015D} and references therein). Thus, rays can be easily derived from the dispersion relation by differentiating it with respect to the wave vector, $\vec{q}$, assuming that $\omega=\omega(\vec{q})$. In order to consider ferrofluids as optical diodes in terms of rays, one should, first of all, find a way to eliminate or disregard solutions to the Fresnel equation that are symmetrical for opposite directions, as it is the case of Eq.~(\ref{vphaseo}). A natural way to do this would be when its associated group velocity is zero or very small. It can be readily checked that this is exactly the case whenever $\epsilon_{33}\approx 0$. Henceforward we will only work within this condition. It is important to stress that $v_{1}^{+}$, the only solution of physical interest here, does not depend on $\epsilon_{33}$ and thus having the latter null only makes the $v_0^{+}$ solution disappear. It is simple to show that the group velocity associated with $v_{1}^{+}$ is
\begin{equation}
\frac{\partial \omega}{\partial \vec{q}}\doteq \vec{u}= \frac{v_1^{+}}{\epsilon_{11}(v_{1}^{+})^2+1}\left(2\hat{q}-\frac{\partial \epsilon_{11}}{\partial B}Ev_1^{+}\hat{x}\right)\label{grouponeway},
\end{equation}
where it has been considered that in the frequency range of interest $\epsilon_{11}$ does not vary significantly with $\omega$. In order to render simple the physical aspects of Eq. (\ref{grouponeway}), let us analyze the situation $\hat{q}$ is in the $xz$-plane. For this case, $\hat{q}_z=\cos\theta$ and $\hat{q}_x=\sin\theta$. When the wave propagates in the plane perpendicular to the magnetic field, similar conclusions to the ones to be drawn will arise. One should simply exchange $\hat{q}_z$ by $\hat{q}_y$ in the subsequent analysis. When the wave is in the plane defined by the external fields, unidirectional propagation of rays does not take place (since $\hat{q}_x=0$ in this case and therefore $v_{1}^+$ is isotropic).

When $\hat{q}$ belongs to the $xz$-plane, one sees from Eq.~(\ref{grouponeway}) that the same happens to the group velocity and its direction is determined by
\begin{equation}
\tan \varphi= \frac{u_x}{u_z}=\tan\theta -\frac{1}{2\cos\theta}\frac{\partial \epsilon_{11}}{\partial B}Ev_{1}^{+}\label{varphi}.
\end{equation}
Observe from Eq. (\ref{grouponeway}) [Eq. (\ref{varphi})] that the $x$-axis is a symmetry axis for the group velocities.

Let us first qualitatively investigate the properties that the group velocity should have. As it can be easily deduced from Eq. (\ref{grouponeway}), once Eq. (\ref{Eeq}) is fulfilled, for $-\pi/2~<~\theta~<~\pi/2$ we have that $\tan\varphi$ admits a critical point at
\begin{equation}
\sin\theta_c=\frac{{\cal A}^2-4\epsilon_{11}}{2{\cal A}^2}\label{thetac},
\end{equation}
where we have defined ${\cal A}=-E(\partial \epsilon_{11}/\partial B)$ only for notational convenience. The associated critical group velocity angle is
\begin{equation}
\tan\varphi_{c}=\sqrt[]{\frac{3{\cal A}^2-4\epsilon_{11}}{{\cal A}^2 + 4\epsilon_{11}}}\label{varphic}.
\end{equation}
From the fact that rays are symmetric with respect to the $x$-axis, $\varphi_c$ actually defines the region in the $xz$-plane where they could propagate. From Eq. (\ref{varphic}) we have that the maximum angle for $\varphi_c$ is $\pi/3$, obtained when $|{\cal A}|\rightarrow\infty$. Nevertheless, this does not correspond to a physical scenario, for much before that the medium should lose its dielectric properties (dielectric breakdown \cite{1988PhRvB..37.2785B}), completely modifying the above reasoning.

In the context of our analysis, one-way propagation of light only takes places when Eq. (\ref{varphic}) has a real solution, which is equivalent to having that $u_x$ in Eq. (\ref{grouponeway}) does not change sign. Thus,
\begin{equation}
\left(\frac{\partial \epsilon_{11}}{\partial B}E\right)^2 \geq \frac{4\epsilon_{11}}{3}\label{Eeq}.
\end{equation}
Since the electric field is still a free parameter, we will assume that the one present in the ferrofluid satisfies the above equation. Estimates of its magnitude will be given in the sequel, when we propose a realization of our analysis.
We turn now to the dependence of $\epsilon_{11}$ on the magnetic field. We will consider that the volume fraction of the inclusions aligned to the magnetic field, giving rise to a wire medium, is \citep{2015arXiv150803019S}
\begin{equation}
\alpha(H) = \alpha_{\infty} \tanh\left(\frac{\bar{\mu} H}{kT} \right)\doteq \alpha_{\infty} \tanh x\label{alpha},
\end{equation}
where the total volume fraction of the inclusions satisfies the constraint $\alpha_{\infty}\ll 1$,
$k$ is the Boltzmann constant ($\approx 1.38\times 10^{-16}$ erg K$^{-1}$) and $\bar{\mu}$ is the intrinsic magnetic moment of the inclusions, assumed to be identical. From the above equation and Eq.~(\ref{e11typ}) it ensues that
\begin{equation}
\frac{\partial \epsilon_{11}}{\partial B}=- \frac{\epsilon_h\alpha_{\infty}}{H}\frac{x}{\cosh^2x}\label{e11partialB}.
\end{equation}
Note that $\partial \epsilon_{11}/\partial B$ is finite in the limit of zero magnetic field. This means that nonlinear effects in magnetic fluids are present even for small applied magnetic fields, which is somewhat counter-intuitive.  Equation (\ref{e11partialB}) has a maximum value, given $\epsilon_h$ and $H$, at the solution of $2x_c\tanh x_c =1$, which is $x_c\approx 0.772$, yielding $x_c/\cosh^2x_c\approx 0.45$.
From Eqs. (\ref{e11typ}), (\ref{Eeq}) and (\ref{alpha}), we obtain that the absolute minimum electric field magnitude that leads to one-way propagation of light is
\begin{equation}
E_{min}\approx 2.566\frac{H}{\alpha_{\infty}\sqrt{\epsilon_{11}}}\label{Emin}.
\end{equation}
Therefore, magnetic liquids could only behave as optical diodes if their dielectric strengths are larger than Eq.~ (\ref{Emin}).

The only physically reasonable situation for our analysis is to consider temperatures above the freezing and below the evaporating points of the host medium. In this regard, we will work with room temperatures ($T=300$ K). For dealing with the smallest possible electric fields, from our previous discussion, one should select magnetic fields of the order of $H=x_c kT/\bar{\mu}$. For ordinary situations ($\bar{\mu}\approx 10^4\mu_B, \mu_B=9.274\times 10^{-21}$ erg/G, the Bohr magneton) [corresponding to nanospheres of around $10$ nm diameter], we have that $H\approx 300$ G, which, from Eq. (\ref{Emin}), would lead to $E_{min}\approx 10^4$ statVolt/cm for $\alpha_{\infty}\approx 10^{-2}$ and $\epsilon_{11}$ of the order of unity. This field is usually above the dielectric breakdown of ordinary liquids, which is of the order of $10^3$ statVolt/cm ($\sim 10^7$ V/m) \citep{kaiser2004electromagnetic}. A solution to this problem is to increase $\bar{\mu}$. If one assumes now that $\bar{\mu}\approx 10^6\mu_B$, then the magnetic field will be of the order of unity, so that the minimum electric field will be smaller than $10^3$ statVolt/cm for the same $\alpha_{\infty}$ and $\epsilon_{h}$ as previously. Therefore, optical diodes made of magnetic fluids could only be realized for nanoparticles with $10^5-10^6$ atoms. This is possible in principle, since they should have diameters around $30$ nm (this could be attained, for instance, with the thermal vaporization method \citep{2005RuCRv..74..489G}), which is smaller than the critical diameter where they lose their single magnetic domain geometry (around $100$ nm \citep{issa2013magnetic,2005RuCRv..74..489G}).

Let us now consider the issue of losses. From Eqs. (\ref{e11typ}) and (\ref{e33typ}) we see that $Im(\epsilon_{11})$ is negligible for the frequency range that we are interested in. For $\epsilon_{33}$ we have that $Im(\epsilon_{33})=\alpha Im(\epsilon_i)$, which is also small even for lossy metals due to $\alpha \ll 1$. Since one-way propagation of light is only related to $v_1^{+}$, one sees that metal losses do not play any role in the effect. They actually reinforce the disregard of the $v_0^{+}$ rays. Therefore, the effect we propose is supposed to have a long propagation length for the rays associated with $v_1^+$.

Kerr effects do not play any role in ordinary magnetic media. The reason is as follows. Such effects rise fundamentally from the dependence of the host permittivity on the electric field, generically written in the form $\bar{\epsilon}_{h}=\epsilon_h + \sqrt[]{\epsilon_h}K\lambda E^2$, where $K$ is the Kerr constant, which for ordinary liquids (transformer oils, for instance) is of the order of $10^{-8}$ cm/(statVolt)$^2$ \citep{2015AIPA....5i7207Y}, and $\lambda$ the wavelength of the propagating waves. One can show that it modifies Eq. (\ref{disprelex}) by a term proportional to $K \lambda E^2$. Thus, corrections to Eq. (\ref{disprelex}) due to the Kerr effect result in a term of the order of $10^{-8}$ for wavelengths of some micrometers, permittivities of the order of unity and electric fields up to the dielectric breakdown. This is small and hence can be disregarded.

Let us now consider a realization of our analysis. For specificity, let us consider  the transformer oil TECHNOL US 4000 as a host medium ($\epsilon_{h}= 2.15$) \citep{2004JMMM..272.2377K} and magnetite Fe$_3$O$_4$ nanoparticles. Analysis with kerosene as the host medium ($\epsilon_h\approx 2.2$) in the presence of cobalt nanoparticles \citep{2014NatSR...4E5706S} would also lead to the same qualitative conclusions as the ones that ensue, given that its dielectric strength is similar to that of transformer oils \citep{kaiser2004electromagnetic}.

Motivated by experimental analysis, let us investigate the particular situation where $\alpha_{\infty}=0.02$ and $H\approx 1$~G. In this case for nanoparticles of $10$ nm, it is known that the breakdown electric field varies from $6$ MV/m to approximately $10$ MV/m when the distance of the electrodes varies from $1$ mm to $0.1$ mm, respectively \cite{2004JMMM..272.2377K}. We assume that similar aspects hold true when the diameters of the nanoparticles are increased to around $30$ nm.
Given that we are working with $T=300$ K and $\bar{\mu}\approx 10^6$$\mu_B$, we have $x\approx 0.22$, which effectively leads to $|E|\gtrsim 185$ StatVolt/cm [see Eqs. (\ref{e11partialB}) and (\ref{Eeq})]. Thus, let us choose $|E|= 190$ StatVolt/cm ($\approx 5.7$ MV/m), related to ${\cal A}= -1.82$ ($E<0$) \footnote{Note that from Eq. (\ref{grouponeway}) the sign of $E$ basically influences the direction rays are allowed to propagate. We have chosen $E<0$ such that the projection of rays along the $z$-direction is only positive (it only points upwards); see Fig. \ref{schematics}. For $E>0$ their $z$-projection would always be negative.}. This value is small enough when compared to the upper breakdown field given previously, which means that the ordinary medium parameters are in principle reliable for electrode distances of the order of some tenths of millimeters.

\begin{figure}[!htb]
 \centering
 \includegraphics[width=\columnwidth]{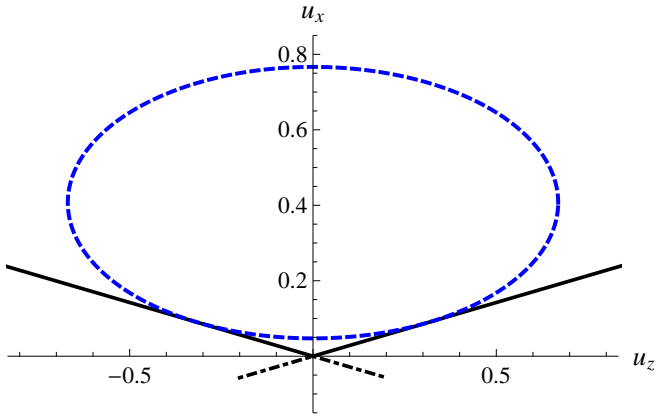}
\caption{{\small\sf (color online). Ray surface associated with Eq. (\ref{grouponeway}) for some transformer oils where $\epsilon_{h}=2.15$, ${\cal A}=-1.82$ and $\epsilon_{33}\approx 0$. The thick straight lines, defined by Eq. (\ref{varphic}), encompass the region of the $xz$-plane where rays can be found. For the parameters chosen, $\varphi_c\approx 16$ degrees.
For each group velocity direction  within the aforementioned region, related only to $v_1^+$, there are actually two associated wave directions.
}}\label{fig1}
\end{figure}

Figure \ref{fig1} depicts the ray surfaces for the aforementioned magnetic liquid. We stress that there is no birefringence involved in our model, for it only deals with one solution to the Fresnel equation, $v_1^+$. Figure \ref{fig1} is to be interpreted in the following way. When one considers waves in all directions of the $xz$-plane, their associated rays (wave packets) are confined in a region of this plane. Therefore, different wave directions must be related to the same ray direction. This is a direct consequence of Eq. (\ref{Eeq}), that guarantees that $u_x$ does not change sign.

For the specific example of Fig. \ref{fig1}, one can show that for fields as high as $11$ MV/m superluminal velocities do not arise. Note that experimental results for $10$ nm nanoparticles suggest that superluminal velocities would only be related to fields beyond breakdown, thus indicating their nonexistence in our model.

Finally, given that ordinary magnetic fluids are colloids, one might wonder what happens when the sizes of the nanoparticles are increased. Actually, it turns out that the drift velocity of a nanoparticle is independent of its mass when gravitational forces are at play \citep{feynman2013feynman}, which means that sedimentation for particles of different masses processes similarly when equilibrium is not attained. Concerning diffusion (due to Brownian motion), at thermodynamic equilibrium the diffusion coefficient depends on the inverse of the mass of the nanoparticles \citep{feynman2013feynman}. This simply means that one should wait for a longer time for heavier particles distribute themselves inside the fluid.

\subsection{Polarization}
The polarizations ($e_i$) for the waves associated with Eqs. (\ref{vphaseo}) and (\ref{vphaseex}) can be found by simply solving its corresponding eigeinvalue equation related to the Fresnel equation (see \citep{2014PhRvA..89d3822D} for the general formalism).
For the case $\hat{q}$ belongs to the $xz$-plane and $\epsilon_{33}\approx 0$ one can show that the polarization related to $v_1^{+}$ is ($b$ is a constant)
\begin{equation}
\vec{e}_{v_1^{+}}=b\hat{y}\label{ev1},
\end{equation}
which is always collinear to the external electric field. Since this is so, the approximation that the resultant electric field is the external one is an outstanding approximation. Therefore, one-way propagation of light only occurs for waves whose polarization is parallel to the electric field and is unaffected by its aspects.

\section{Summary}

In this work we have shown that some fluids with magnetic nanoparticles of $30$ nm diameter (with magnetic moments of approximately $10^6\mu_B$) in concentrations around $2\%$ or larger may behave as optical diodes for the propagation of light rays in the presence of small DC magnetic fields and DC electric fields of around half of their dielectric breakdown. The effect only takes place when $\epsilon_{33}\approx 0$, implying that $v_0^+$ rays (extraordinary rays) are suppressed, leaving only $v_1^+$ rays (generalization of ordinary rays in linear media), which have negligible losses and are polarized in the direction of the external electric field. Experimental results regarding dielectric breakdown of magnetic fluids with $10$ nm nanoparticles suggest that superluminosity does not arise in our model, indicating thus its physical reasonableness.

\acknowledgements
We thank the anonymous referee for his/her valuable suggestions that helped us improve this manuscript. J.P.P. is grateful to CNPq-Conselho Nacional de Desenvolvimento Cient\'ifico e Tecnol\'ogico of the Brazilian government within the postdoctoral program ``Science without Borders'' for the financial support. Thanks also goes to FAPESP-Funda\c c\~ao de Amparo \`a Pesquisa do Estado de S\~ao Paulo, for partial support. We are indebted to Prof. Dr. Michael Farle for useful comments on properties of magnetic nanoparticles. This work was supported in part by NSF grant DMR-1104676.

%

\end{document}